\documentclass[twocolumn,floatfix]{revtex4}

\usepackage{graphicx,bm,psfrag, amsmath, color}

\usepackage[dvipsnames]{xcolor}
\usepackage[utf8]{inputenc}

\newcommand{\beq}{\begin{equation}}
\newcommand{\eeq}{\end{equation}}
\newcommand{\beqa}{\begin{eqnarray}}
\newcommand{\eeqa}{\end{eqnarray}}
\newcommand{\ba}{\begin{array}}
\newcommand{\ea}{\end{array}}

\usepackage{youngtab}

\newcommand{\be}{\begin{equation}}
\newcommand{\ee}{\end{equation}}
\newcommand{\bea}{\begin{eqnarray}}
\newcommand{\eea}{\end{eqnarray}}

\newcommand{\Tc}{T_{\mathrm c}}
\newcommand{\Tosc}{T_{\mathrm osc}}
\newcommand{\Tr}{T_{\mathrm r}}

\newcommand{\Lc}{\Lambda_{\mathrm c}}
\newcommand{\fc}{f_{k}}

\usepackage[english]{babel}
\usepackage[T1]{fontenc}
\usepackage[colorinlistoftodos]{todonotes}
\usepackage[colorlinks=true, allcolors=blue]{hyperref}

\begin{document}

\title{
All-in-one Relaxion, \\a unified  solution to five BSM puzzles 
}
\author{R.S.~Gupta, J.Y.~Reiness and M.~Spannowsky}
\affiliation{Institute for Particle Physics Phenomenology,
Durham University, South Road, Durham, DH1 3LE}

\date{\today}

\begin{abstract}
We present a unified relaxion solution to  the five major  outstanding issues in particle physics: the hierarchy problem, dark matter,  matter-antimatter asymmetry,  neutrino masses and the strong CP problem. The only additional field content in our construction with respect to standard relaxion models is an up-type vector-like  fermion pair and three right-handed neutrinos charged under the relaxion shift symmetry. The observed  dark matter abundance is generated automatically by oscillations of the relaxion field that begin once it  is  misaligned from its original stopping point after reheating. The  matter-antimatter asymmetry arises from spontaneous baryogenesis induced by the CPT violation due to the rolling of the relaxion  after reheating.  The CPT violation is communicated to the baryons and leptons via an operator, $\partial_\mu \phi J^\mu$, where $J^\mu$ consists of right-handed neutrino currents arising naturally from a simple neutrino mass model. Finally, the strong CP problem is solved via the Nelson-Barr mechanism, i.e. by imposing CP as a symmetry of the Lagrangian that is broken only spontaneously by the relaxion. The CP breaking is such that although an ${\cal O}(1)$ strong CKM phase is generated, the induced strong CP phase is much smaller, i.e., within experimental bounds.
\end{abstract}

\maketitle

\section{Introduction}
In recent times particle physics research has been driven to a large extent  by the expectation of physics beyond the Standard Model (BSM) at the TeV scale. While there are many theoretical and observational reasons to extend the Standard Model (SM) -- such as the hierarchy problem, dark matter,  matter-antimatter asymmetry,  neutrino masses and the strong CP problem -- only the first of these issues is believed to necessarily require TeV scale new physics. 

In fact if  the hierarchy problem is ignored and  new physics scales much beyond the TeV scale are allowed, the other issues can be solved by very minimal extensions of the SM.   For instance, in the so-called `SMASH' model of Ref.~\cite{Ballesteros:2016euj,Ballesteros:2016xej}, a simultaneous solution  to neutrino masses,  baryogenesis, the strong CP problem, dark matter and a model for inflation was achieved with the addition of only  three right-handed SM-singlet neutrinos, a new vector-like color triplet fermion and a complex SM singlet scalar. In another set of papers (the flaxion/axiflavon constructions of Ref.~\cite{Ema:2016ops,Calibbi:2016hwq}), the Peccei-Quinn   solution~\cite{Peccei:1977hh} to the strong CP and dark matter puzzles was unified with the Froggatt-Nielsen solution~\cite{Froggatt:1978nt} to the SM flavour puzzle by identifying the Froggatt-Nielsen  and Peccei-Quinn  symmetries. Furthermore in Ref.~\cite{Ema:2016ops}, it was also shown that the radial component of the flavon field in such models can act as the  inflaton and the baryon asymmetry can be explained by leptogenesis.

It is arguably much more challenging to find an explanation (apart from tuning or anthropics) for a light Higgs mass with a high new physics scale. While  conventional wisdom says this is impossible, the recently-proposed  cosmological relaxation (or relaxion) models~\cite{Graham:2015cka} aim to find just such an explanation.  In these models the rolling of the so-called relaxion field during inflation leads to a scanning of the the Higgs mass squared  from   positive to negative values. Once the Higgs mass squared becomes negative it triggers a backreaction potential that stops the scanning soon after, at a value much smaller than the cut-off, thus dynamically solving the hierarchy problem.

Building on previous  work~\cite{Davidi:2017gir,Davidi:2018sii,Abel:2018fqg}, we show in this paper that the relaxion construction already has ingredients that can provide solutions to multiple BSM puzzles  in addition to the hierarchy problem.  
Our starting point is Ref.~\cite{Abel:2018fqg}, in which it was shown that Spontaneous Baryogenesis~\cite{Cohen:1987vi,Cohen:1988kt}  can be successfully achieved in scenarios where the relaxion rolls after reheating (due to the vanishing of the backreaction at high temperatures).
The spontaneous CPT violation due to the relaxion  rolling   biases thermal equilibrium such that sphalerons processes lead to a net generation of $B+L$ number. The CPT breaking is communicated to the baryons/leptons via an operator, ${\cal O}_{SB}=\partial_\mu  \phi J^\mu$, where $J^\mu$ is a current containing the $B+L$ current.   Once the temperature drops sufficiently the backreaction potential reappears and the relaxion, being misaligned from the original minima, starts oscillating~\cite{Banerjee:2018xmn}. Remarkably, in the finite parameter space   where baryon asymmetry can be successfully explained, relaxion oscillations can also explain the observed dark matter abundance~\cite{Abel:2018fqg}. 

 In this paper we improve the above spontaneous relaxion baryogenesis (SRB) set-up of Ref.~\cite{Abel:2018fqg}   by clarifying the origin of its two non-renormalisable parts, namely the operator ${\cal O}_{SB}$ and the rolling potential. The first improvement is a much needed explanation for the operator ${\cal O}_{SB}$. In this work, the current $J^\mu$  consists of a combination of right-handed neutrino currents that arises naturally from a neutrino mass model; the operator  ${\cal O}_{SB}$ would thus be practically unconstrained, leading to an opening up of the parameter space with respect to Ref.~\cite{Abel:2018fqg}. 
 
 Finally, we provide a renormalisable sector that can radiatively generate the relaxion rolling potential. Borrowing from the Nelson-Barr   relaxion construction of Ref.~\cite{Davidi:2017gir}, we present a model that  generates the rolling potential while also solving the strong CP problem via the Nelson-Barr  mechanism~\cite{Nelson:1983zb, Barr:1984qx}. Following the basic philosophy of Nelson-Barr models,  CP is assumed to be an exact symmetry in the UV and is only broken spontaneously by the relaxion  in a way that does not generate a strong CP phase greater than allowed limits. 
 On the other hand, the   phase of  the relaxion at its stopping point gets mapped to the CKM phase.

 Thus, with a well-motivated completion of the minimal SRB  scenario of Ref.~\cite{Abel:2018fqg} we achieve a unified solution to five  BSM puzzles, namely the hierarchy problem, dark matter,  matter-antimatter asymmetry,  neutrino masses and the strong CP problem.  We describe our set-up in detail in the following sections.

%$(\psi, \psi^c)$ 

\section{Review and basic set-up}

In relaxion models, the Higgs mass squared parameter, $\mu^2$, is promoted to a dynamical quantity $\mu^2(\phi)$, which varies due to its couplings to the new relaxion  field, $\phi$,
\begin{equation}
\label{eq:roll}
V_{\mathrm{roll}}=\mu^2(\phi)H^\dagger H+\lambda_H(H^\dagger H)^2 -r_{\mathrm{roll}}^2 M^4\cos\dfrac{\phi}{F},
\end{equation}
with,
\begin{equation}
\label{eq:mu2}
\mu^2(\phi)=\kappa M^2-M^2\cos\dfrac{\phi}{F}.
\end{equation}
Here, $H$ is the  SM Higgs doublet, $\lambda_H$ is its quartic coupling, and $M$ is the UV cut-off of the Higgs effective theory \footnote{The periodicity $F$ is large such that the cosine is locally linear in $\phi$.  By expanding the cosine terms about an arbitrary point,  the polynomial terms of the original work~\cite{Graham:2015cka} can be recovered with $g \sim M^2/F$.}. The coefficient $r_{\mathrm{roll}}$ is model-dependent and $\kappa \lesssim 1$. During inflation the relaxion field slow-rolls due to the last term in Eq.~\ref{eq:roll} such that the value  $\mu^2$ parameter is slowly scanned starting from positive values. Starting from a relaxion field value $\phi < \phi_c = -|F\cos^{-1}\kappa|$, the relaxion field slow-rolls down the  slope, increasing the value of $\phi$ and decreasing the value of $|\mu^2|$. After crossing the point $\phi = \phi_c$, $\mu ^2 $ becomes negative prompting electroweak symmetry breaking. This in turn activates the backreaction potential which induces periodic `wiggles' on top of the linear envelope,
\begin{equation}
\label{eq:vbr}
V_{\mathrm{br}}=\Lambda_c^4\cos \dfrac{\phi}{\fc},
\end{equation}
 were $\Lambda_c^4= m^n v(\phi)^{4-n}$ is an increasing function of the Higgs vacuum expectation value (VEV). These wiggles cause the relaxion field to come to a halt at $\phi=\phi_0$ satisfying,
\begin{equation}
\label{eq:halt}
V_{\mathrm{roll}}'(\phi)+V_{\mathrm{br}}'(\phi)=0 \quad
\Rightarrow \quad
\dfrac{M}{\Lambda_c (v)}\sim \left( \dfrac{F}{r_{\mathrm{roll}}^2 \fc}  \right)^{1/4},
\end{equation}
where we have taken  $\sin(\phi_0/\fc)\sim \sin(\phi_0/F)\sim\mathcal{O}(1)$. Hence, if $F/f_k\gg 1$, a large hierarchy can be achieved between the Higgs VEV, $v$, and the cut-off  $M$. 
%\JYR{More discussions straight out of previous paper}
%
%
%
%
%
%
%\subsection{The Timeline \& Cosmology}
%
%Order in which things happen and different scales involved. \JYR{Same as previous paper}
%
%As explained in \cite{Graham:2015cka}, there are two conditions on the Hubble scale of inflation $H_i$. A brief summary:
%\begin{enumerate}
%	\item Vacuum energy $>$ vacuum energy change along $\phi$ potential, $H_i^2\gtrsim V(\phi)/M_P$, and thus,
%	\begin{equation}
%	H_i\gtrsim \dfrac{r_{\mathrm{roll}}\Lambda_H^2}{M_P},
%	\end{equation} 
%	since $V\sim r_{\mathrm{roll}}^2\Lambda_H^4$ at inflation.
%
%	\item Classical rolling dominates over quantum fluctuations, $H_i<V'/H_i^2$, such that,
%	\begin{equation}
%	H_i < \left(  \dfrac{r_{\mathrm{roll}}^2\Lambda_H^4}{F} \right)^{1/3}.
%	\end{equation}
%	
%\end{enumerate}
%There is a third condition concerning the generation of the cosine potential in the backreaction. We leave discussion of the specifics of the backreaction to Section \ref{sec:clockk}.

As discussed in Ref.~\cite{Graham:2015cka}, the cut-off, $M$, cannot be raised to an arbitrarily high value  because of the following cosmological requirements: (1) the relaxion dynamics should be classical, i.e., quantum fluctuations should be sub-dominant, which implies that $H_I<V'/H_I^2$, $H_I$ being the Hubble scale during inflation and (2) the relaxion vacuum energy should not drive inflation and should thus be negligible compared to the total vacuum energy, i.e. $M^4<H^2 M^2_{Pl}$. Together these conditions give the following upper bound on the cut-off (where we have used Eq. \ref{eq:halt}),
\begin{equation}
\label{eq:relcos}
M \lesssim \left( \dfrac{M_P}{r_\mathrm{roll}} \right) ^{1/2}
\left( \dfrac{\Lambda_c^4}{f_k} \right)^{1/6}.
\end{equation}

\begin{figure}
\noindent \begin{centering}
\includegraphics[scale=0.3]{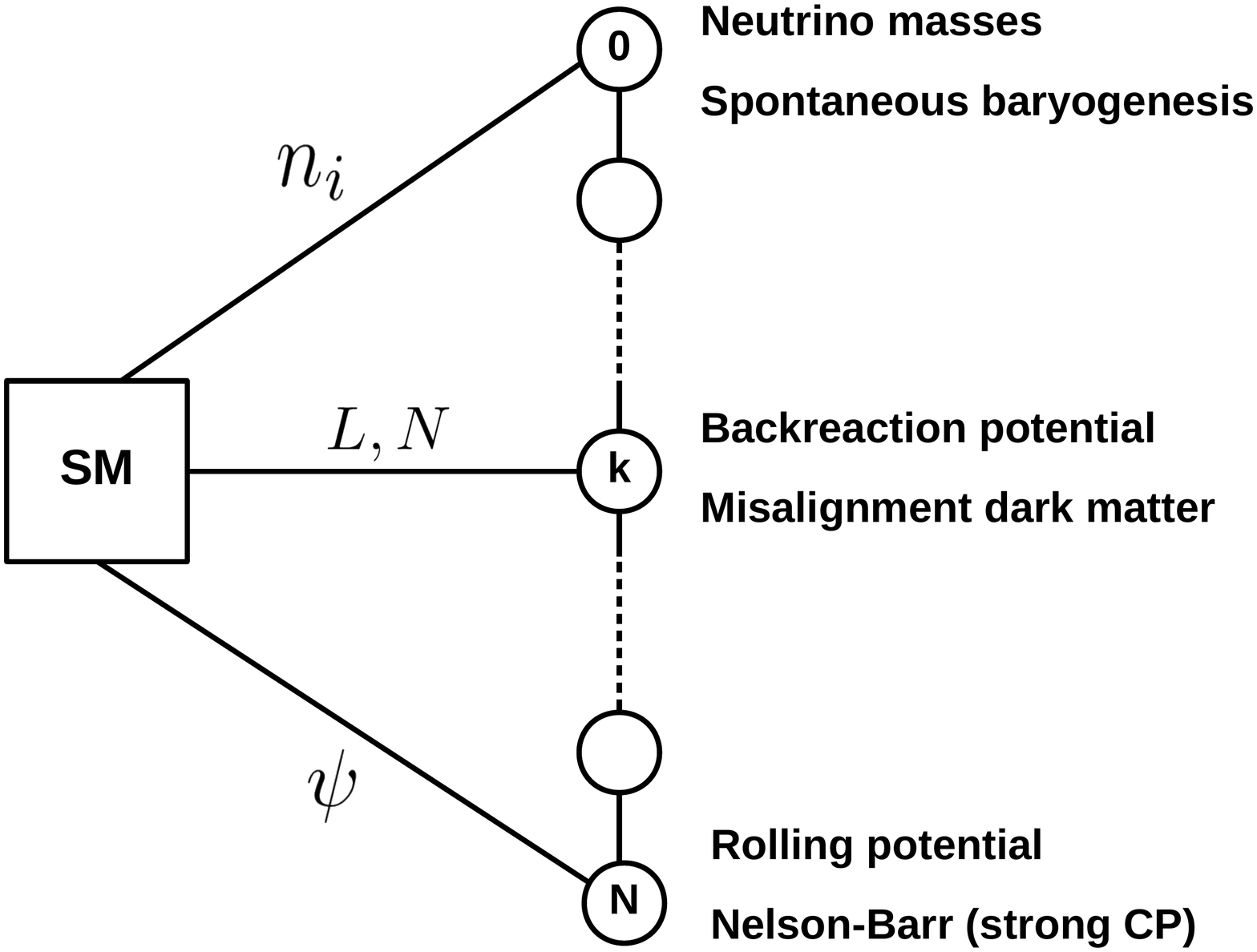}
\par\end{centering}
\protect\caption{Schematic representation of our set-up with the vertical line representing the clockwork system. The  three right-handed neutrinos at the $0$th site, $n_i$, couple to the SM in the usual way, generating neutrino masses; they also provide the current in the all-important operator for spontaneous baryogenesis, ${\cal O}_{SB}=\partial_\mu \phi J^\mu$. At the $k$th site we introduce a new strong sector which couples to SM via its fermionic matter content,  $(L,L^c,N,N^c)$. This sector generates  the backreaction wiggles and  relaxion oscillations inside these wiggles generate the observed dark matter abundance.   Finally, at site $N$ there is a  Nelson-Barr sector that radiatively generates the rolling potential while also providing a   solution to the strong CP problem. This sector couples to the SM up sector via a new vector-like quark pair $(\psi,\psi^c)$.}
\label{fig:diag}
\end{figure}
Let us now discuss what happens after inflation. First note that for the backreaction  sector we will adopt the  non-QCD model of Ref.~\cite{Graham:2015cka} where $\phi$ is the axion of a new strong sector. This sector couples to SM via vector-like electroweak doublets and singlets   $(L,L^c)$ and $(N,N^c)$ that are charged under the new strong group and are given Yukawa couplings to the Higgs. If the reheating temperature is greater than the critical temperature of the chiral phase transition of the new sector, i.e. if $\Tr> \Tc \sim \sqrt{4 \pi} f_{\pi'}$, the wiggles disappear and the relaxion starts rolling again. Here $f_{\pi'}$ is the analog of the `pion decay constant' of the strong sector. When the universe cools  below  the temperature $\Tc$ again,  the backreaction potential reappears and the rolling eventually stops provided,
 \begin{equation}
m_\phi ~\lesssim~ 5 H(\Tc) ~.
\label{eq:slowroll}
 \end{equation}
This    condition  is obtained by demanding  that the relaxion does not pick up enough kinetic energy to overshoot the barriers once the backreaction potential reappears~\cite{Kawasaki:2011pd,Banerjee:2018xmn,Choi:2016kke}.  We note for future reference that the backreaction scale $\Lc$ depends not just on $f_{\pi'}$ but also on other technically-natural couplings~\cite{Graham:2015cka}, such that  $\Lc$ and $f_{\pi'}\sim \Tc$ are two independent scales.  Also note that in this model we must have  $\Lc^4 \lesssim 16 \pi^2 v^4$  so that Higgs-independent contributions are sub-dominant compared to Higgs-dependant ones.

It is this second phase of rolling that can lead to a generation of both the observed dark matter abundance as well as  the baryon asymmetry.  The explanation for dark matter requires no additional ingredient. This is due to the fact that during the second phase of rolling, the relaxion gets misaligned from its original stopping point by an angle~\cite{Banerjee:2018xmn},
  \begin{equation}
\Delta \theta~=~\frac{\Delta \phi}{f} ~\simeq~ \frac{1}{20}\left(\frac{m_\phi}{H(\Tc)}\right)^2 \tan \frac{\phi_0}{f}~.
\label{eq:misalignment}
 \end{equation}
 As shown in Ref.~\cite{Banerjee:2018xmn}, this sets  off relaxion oscillations that can give rise to the observed dark matter relic abundance, 
  \begin{equation}
\Omega h^2 ~\simeq~ 3 \Delta \theta^2 \left(\frac{\Lambda_d}{1~{\rm GeV}}\right)^4  \left(\frac{100~{\rm GeV}}{\Tosc }\right)^3~.
\label{eq:dm}
 \end{equation}
 Note that the correct relic density can always be reproduced by choosing an appropriate value of $\tan \frac{\phi_0}{f}$. While there is some room for this in the relaxion mechanism, as the relaxion is spread across multiple vacua at the end of its rolling,  the probability distribution of the relaxion field peaks for ${\cal O}(1)$ values of   $\tan \frac{\phi_0}{f}$~\cite{measure}. Thus the extent to which $\tan \frac{\phi_0}{f}$  deviates from unity can be interpreted as a measure of the tuning required to get the correct relic abundance.

It was shown in Ref.~\cite{Abel:2018fqg} that with just one additional requirement, this second phase of rolling can also explain the observed baryon asymmetry. One additionally requires that some fermions with $B+L$ charge must be charged under the relaxion shift symmetry. This leads to the presence of the operator, $\partial_\mu \phi J^\mu$, where $J^\mu$ contains the $B+L$ current. This operator can generate a chemical potential for $B+L$ violation once the second phase of relaxion rolling results in a  CPT breaking  expectation value for $ \partial_\mu \phi$, giving rise to a baryon asymmetry due to the presence of $(B+L)$-violating sphaleron transitions.

 As we will show later, to generate a baryon asymmetry of the observed size one requires a hierarchy $f \ll \fc$. This and the fact that the relaxion, in any case, requires a large hierarchy between  $\fc$ and its  field excursion during rolling, $\fc \ll F$, are problematic as  pointed out in Ref.~\cite{Gupta:2015uea}. The solution to generating the latter hierarchy is embedding  the relaxion construction in a so-called clockwork model~\cite{Choi:2014rja,Choi:2015fiu,Kaplan:2015fuy}; this can easily be extended to also generate the former hierarchy, giving the overall hierarchy $f \ll \fc \ll F$. In clockwork models there is a system  of interacting complex scalars, $\Phi_i$, all of which get a VEV such that $\langle \Phi_i \rangle = \frac{f}{\sqrt{2}} e^{i \pi_i/f}$. There is an approximate abelian symmetry, U(1)$_i$,  at each site which is spontaneously broken to give rise to a corresponding pseudo-goldstone mode $\pi_i$. Explicit breaking effects give the angular fields,  $\pi_i$,   a mass matrix such that the lightest state is a massless (Goldstone) mode given by, 
 \begin{equation}
\phi \propto \sum_j\dfrac{\pi_j}{3^j}=\pi_0+\dfrac{\pi_1}{3}+...+\dfrac{\pi_N}{3^N}.
\label{flatdir}
\end{equation}
Given that the mixing angle $\langle \pi_k |\phi\rangle \sim 3^{-k}$, any Lagrangian term where the  angular field $\pi_k$ couples with a decay constant $f$ translates to an interaction of $\phi$ with an exponentially-enhanced effective decay constant, $3^k f$, in the mass basis.  Coming back to our set-up, the hierarchy $f \ll f_k \ll F$ can be obtained by having the operator, ${\cal O}_{SB}$, at the $0$-th site, the backreaction sector  at the intermediate $k$-th site and the rolling potential at the $N$th site, such that $f_k=3^k f$ and $F=3^N f$. This is schematically shown in Fig.~\ref{fig:diag}. The rolling and backreaction potentials eventually lift the  flat direction in Eq.~\ref{flatdir}, giving a mass to the relaxion.

While the  SRB set-up reviewed above is an elegant solution to multiple BSM puzzles with only a minimal modification of standard relaxion models, it is incomplete in two respects:  the operator, ${\cal O}_{SB}$,  and the rolling potential are non-renormalisable and introduced in a somewhat ad hoc way. While  ${\cal O}_{SB}$ arises naturally if baryons and/or lepton are charged under the abelian symmetry of which the relaxion is a goldstone boson, charging the SM fermions seems to have no purpose other than generating ${\cal O}_{SB}$. Furthermore, the charge assignments have to be carefully chosen such that they are anomaly free with respect to QCD (a necessary condition to avoid generating a strong CP phase) and preferably also with respect to electromagnetism (to avoid the generation of a $\phi \gamma \gamma$ coupling that rules out most of the parameter space in this set-up~\cite{Abel:2018fqg}). Here we complete the SRB set-up as follows:
\begin{itemize}
\item{Instead of introducing the operator, ${\cal O}_{SB}$,  by hand, we propose a simple neutrino mass model at site $0$, which generates this operator with a current containing only three new right-handed (RH) neutrino fields. The operator arises because the RH fields are charged under the relaxion shift symmetry, which in turn  is identified with a Froggatt-Nielsen symmetry. This also explains the observed smallness of neutrino masses. As only the SM singlet RH neutrinos are charged under it, the relaxion shift symmetry is automatically anomaly-free with respect to both QCD and electromagnetism. }
\item{We show that the rolling potential can be generated by the addition of a single up-type vector-like pair at the site $N$. In addition, this modification can solve the strong CP problem via the Nelson-Barr   mechanism~\cite{Nelson:1983zb, Barr:1984qx}. In  Nelson-Barr models, CP is a good symmetry in the UV and is spontaneously broken at an intermediate scale to generate an ${\cal O}(1)$ CKM phase but a much smaller strong CP phase (within allowed constraints). We borrow the Nelson-Barr relaxion sector from Ref.~\cite{Davidi:2017gir}, where the relaxion phase  upon stopping results in an  ${\cal O}(1)$ CKM phase. }
\end{itemize}

From the point of view of the Nelson-Barr relaxion model, baryogenesis is not straightforward to achieve due to the lack of any explicit CP violation. Spontaneous   baryogenesis is thus an attractive potential feature for the Nelson-Barr relaxion, as it does not require explicit CP violation.

\section{ Neutrino masses and Spontaneous Baryogenesis}
\label{sbg}

\subsubsection{Getting the operator $\mathcal{O}_{SB}$}

At site $0$ we introduce a sector that simultaneously generates small neutrino masses and an operator suitable for spontaneous baryogenesis. We introduce three RH neutrinos, $n_i$, that are charged under the abelian symmetry at the 0-th site of the clockwork chain, U(1)$_0$. We fixed the charge of   $\Phi_0$ to be   $-1$ under this symmetry and take all SM fields to be neutral. The Lagrangian for the couplings of these right-handed  neutrinos is given by,
\begin{equation}
\mathcal{L}_{\mathrm{FN}}\supset y^{ij}_n(\Phi_0/\Lambda_{FN})^{q_{n_j}}{l}_iHn^\dagger_j+M_n^{ij}n_in_j,
\label{neut}
\end{equation}
where $l_i$ are the SM lepton doublets,  $q_{n_j}$ are the abelian charges for the sterile neutrinos and $M_n^{ij}$ are the associated Majorana mass matrix elements. Given that we will eventually use the Nelson-Barr solution to the strong CP problem,  we impose that CP is an exact symmetry of the Lagrangian so that all the couplings above are real. 
Following spontaneous symmetry breaking, the first term becomes,
\begin{equation}
\label{eq:FNphase}
y_n^{ij}(\epsilon_{FN})^{q_{n_j}}e^{i\pi_0q_{n_j}/f}l_iHn^\dagger_j,
\end{equation}
where $\epsilon_{FN}={f}/{\sqrt{2}\Lambda_{FN}}<1$. The effective Yukawa couplings, $Y_n^{ij}=y_n^{ij}(\epsilon_{FN})^{q_{n_j}}$  can thus be exponentially small, which, as we will soon see, provides a natural explanation for neutrino masses with a smaller-than-usual seesaw scale. This is just an implementation of the Froggatt-Nielsen mechanism~\cite{Froggatt:1978nt} to naturally obtain small Yukawa couplings. While the first term in Eq.~\ref{neut} is still non-renormalisable, we note that such operators can be UV completed in standard ways involving chains of vector-like fermions (see for eg. Ref.~\cite{Davidi:2018sii}).

The factor  $e^{i\pi_0q_{n_j}/f}$ in front of the Yukawa couplings can be rotated away by the field redefinition $n_j\to n_j e^{-i\pi_0q_{n_j}/f}$, which, through the redefinition of the kinetic terms for the RH neutrinos, yields the desired operator, $\mathcal{O}_{SB}$, 
\begin{equation}
\dfrac{q_{n_i}}{f}(\partial_\mu\pi_0){n^\dagger_i}\bar{\sigma}^\mu n_i \to \dfrac{q_{n_i}}{f}(\partial_\mu\phi) {n^\dagger_i}\bar{\sigma}^\mu n_i,
\end{equation}
where we ignore an ${\cal O}(1)$ factor corresponding to $\langle \pi_0|\phi\rangle$. 
\subsubsection{Getting Baryon Asymmetry from $\mathcal{O}_{SB}$}

Now we will demonstrate how the presence of $\mathcal{O}_{SB}$ can lead to  spontaneous baryogenesis, an idea developed  in Refs.~\cite{Cohen:1987vi,Cohen:1988kt}. The essential feature of this mechanism is the presence of a rolling  field that breaks CPT. It was shown in Ref.~\cite{Abel:2018fqg} that this role can be played by the rolling of the  relaxion after reheating.

During the second phase of the rolling of the relaxion, the operator $\mathcal{O}_{SB}$ causes equal and opposite shifts in the energy of particles versus antiparticles, implying,
\begin{align}
\mu_i &= - \bar{\mu}_i ~=~ 
 q_{i} \dot{\phi}/f + (B_i-L_i) \mu_{B-L} + Q_i \mu_{Q}+ T_{3i} \mu_{T_3}~,\nonumber
\end{align}
where  $\mu_i$ ($\bar{\mu}_i$) is the chemical potential for (anti-)~particles of the $i$th specie; $q_i$ is its charges under U(1)$_0$ (which is non-zero only for the RH neutrinos); $Q_{i}$ is the electromagnetic charge; $T_{3i}$ is the charge corresponding to the diagonal generator of SU(2)$_L$;  and the chemical potentials $\mu_{Q,T_3,B-L}$  have been introduced to enforce conservation of $Q, T_3$ and $B-L$. In the presence of $(B+L)$-violating sphaleron processes, we find,
\begin{align}
\label{eq:rho}
n_{i} - \bar{n}_i &~=~  f(T,\mu) - f(T,\bar{\mu})\\
&~=~ g_i \mu_i  \frac{T^2}{6},~ g_i \mu_i  \frac{T^2}{3}, \nonumber
\end{align}
for fermions and bosons respectively, where $f(T,\mu)$ is the Fermi-Dirac (Bose-Einstein) distribution for fermions (bosons). We have taken $\mu \ll T$ and the factor $g_i$ denotes the number of degrees of freedom for each species. The quantities  $\mu_{T_3,Q,B-L}$ can be obtained by  imposing $n_{T_3}=n_{Q} = n _{B-L} =0$. For temperatures above the critical temperature for the electroweak phase transition, we obtain the following chemical potentials,
\begin{equation}
\mu_{Q}=-\frac{3}{14}\frac{Q_n \dot \phi}{f}~~\mu_{T_3}=\frac{3}{14}\frac{Q_n \dot \phi}{f}~~\mu_{B_L}=\frac{33}{112}\frac{Q_n \dot \phi}{f},
\end{equation}
taking all the $q_{n_i}=Q_{n}$. We subsequently obtain a baryon number density,
\begin{equation}
n_B=-n_L= g_{\rm SB}\frac{\dot\phi}{f}\frac{T^2}{6},
\label{nbl}
\end{equation}
where $g_{\rm SB}=3~Q_n/4$ and finally for its ratio with the entropy density,
\begin{equation}
\label{eta1}
\eta ~\equiv~ \frac{n_B}{s} ~=~ g_{\rm SB}\frac{\dot\phi}{f}\frac{T^2}{6} \times  \left(\frac{2\pi^2 g_* T^3}{45}\right)^{-1} ~=~ \frac{15}{4\pi^2}\frac{g_{\rm SB}}{g_*} \frac{\dot \phi}{f T}~.
\end{equation}
 The equilibrium distribution changes after electroweak symmetry breaking, when there is no longer a need to conserve $T_3$. This gives $\mu_Q=-(4/11) \mu_{B-L}=-{Q_n \dot \phi}/{12 f}$ and, once again, $g_{\rm SB}=3Q_n/4$. However, species such as the RH fermions,  which are coupled very weakly to the thermal plasma, would not be able to re-equilibrate on the time scale of the electroweak phase transition. The precise value of $g_{\rm SB}$ is thus  hard to compute without considering the full dynamics of the process and may be different from the value obtained above. Given that these  subtleties only lead to an ${\cal O}(1)$ ambiguity in $g_{\rm SB}$, for definiteness we stick to the value derived in Eq.~\ref{nbl}. 
 
  The value of  $\eta$ is frozen at $T=T_{sph}=130$ GeV, the temperature at which the sphaleron processes decouple~\cite{DOnofrio:2014rug}. Requiring that this reproduces the observed baryon asymmetry,  $\eta_0= 8.7 \times 10^{-11}$, we obtain,
\begin{equation}
\dfrac{\fc}{f}=\sqrt{\frac{2}{5}}\dfrac{2\pi^3 g_*^{3/2} T_{sph}^3 \eta_0}{9 g_{SB}m_\phi^2 M_{Pl}}\sim  10^9 \left( \dfrac{m_\phi}{10^{-5}\mathrm{eV}} \right)^{-2},
\label{ratio}
\end{equation}
where we have used $V'\sim 5H\dot{\phi}\sim \Lambda_c^4/\fc$ via Eq.~\ref{eq:slowroll}. To obtain the numerical value above we have taken the value of $g_{\rm SB}$ in Eq.~\ref{nbl} with $Q_n=6$, where the latter choice will be justified in the next subsection. It is crucial that the relaxion keeps rolling with a non-zero $\dot \phi$ when the value of $\eta$ gets frozen at $T=T_{sph}$. To ensure this, we require that the critical temperature for the phase transition of the strong sector, $\Tc$, is lower than $T_{sph}$.

Note that in the above analysis we have assumed that the RH neutrinos are relativistic and in equilibrium at the temperatures relevant for the calculation above. First of all, this requires that the seesaw scale,
\begin{equation}
 M_n\lesssim T_{sph},
 \label{maj}
 \end{equation}
  where $M_n^{ij}=M_n$. The second requirement -- that the RH neutrinos are in equilibrium -- implies that the interaction rate of $n_i$ with SM particles satisfies, 
\begin{equation}
\Gamma(n)>H(T_{sph}).
\end{equation}
As discussed in detail in Ref.~\cite{Dick:1999je}, taking into account all the processes that contribute to the  equilibration of the RH neutrinos, we dimensionally expect, $\Gamma(n) \sim g^2Y_N^2T$, where $g$ is the weak coupling and $Y_n^{ij}\sim Y_n$. Requiring $\Gamma>H(T_{sph})$ gives,
 \begin{equation}
 Y_n\gtrsim 10^{-8}.
 \label{equib}
 \end{equation}

\subsubsection{Neutrino masses}

The  Lagrangian in Eq.~\ref{neut} generates masses for the SM neutrinos,
\begin{equation}
m_\nu \sim Y_n^2v^2/M_n\lesssim 0.1\mathrm{~eV}.
\end{equation}
Given that spontaneous baryogenesis demands $M_n\lesssim  T_{sph}$, we require  small effective Yukawas $Y_n\lesssim 10^{-6}$, which can be naturally obtained due to the Froggatt-Nielsen mechanism,  as explained above. Note that without the Majorana mass term the neutrinos would only  have a Dirac mass, which would require a much smaller Yukawa coupling.  Our explanation for  the smallness of neutrino masses thus utilises a combination of the Froggatt-Nielsen and seesaw mechanisms.

 Eq.~\ref{maj} and Eq.~\ref{equib} imply that  our model requires a finite range for both the Yukawa coupling and Majorana mass scale, 
\begin{eqnarray}
10^{-8} \lesssim &Y_n& \lesssim 10^{-6}\nonumber\\
30 \mathrm{~MeV}\lesssim &M_n&\lesssim T_{sph}.
\end{eqnarray}
While  sterile neutrinos with masses   below 500 MeV  are in tension with big-bang nucleosynthesis (BBN),  for masses around a few GeV they might not be too far from the reach of future experiments such as SHiP~\cite{Alekhin:2015byh}. For $\epsilon_{FN}=0.1$ the above range of the Yukawa couplings can be obtained for $6\leq Q_n\leq 8$, where we have taken all $q_{n_i}=Q_{n}$. We have chosen equal charges for all the RH neutrinos because our objective is just to illustrate how the smallness of the neutrino mass scale can be explained. While this will result in an anarchic neutrino mass matrix (for an anarchic $y^{ij}_n$) more structure in the masses and mixings can be obtained, for instance, by also charging the lepton doublets under the Froggatt-Nielsen symmetry (for an example  see Ref.~\cite{Ema:2016ops}). Finally note that as all the couplings in Eq.~\ref{eq:FNphase} are real, a concrete prediction of our model is a CP-conserving phase in the neutrino mass matrix which is still well within the range allowed by neutrino experiments ~\cite{nfit}.

\section{Nelson-Barr sector and the rolling potential}
\label{nbr}

\subsubsection{Generating the Rolling Potential}

To generate the rolling potential the abelian symmetry  at the last site of the clockwork, U(1)$_N$, needs to be broken. In Ref.~\cite{Davidi:2017gir} it was shown that this can be achieved with  a minimal modification of the SM up sector, namely the addition of a vector-like pair, $(\psi, \psi^c)$, where $\psi$ has the same quantum numbers as an up-type singlet. Furthermore, as shown in   Ref.~\cite{Bento:1991ez},  the same modification can also give a Nelson-Barr solution~\cite{Nelson:1983zb, Barr:1984qx} to the strong CP problem, provided we impose an additional $Z_2$ symmetry. The Lagrangian terms for the relevant interactions are,
\begin{align}
\mathcal{L}_{\mathrm{NB}}&=Y^u_{ij} Q\tilde{H} u^c+y^i_\psi\psi \Phi_N u^c_i + \tilde{y}^i_\psi \psi  \Phi_N^*u^c_i \nonumber\\&+\mu_\psi\psi\psi^c+h.c.
\label{eq:nb}
\end{align}
where the  $\psi, \psi^c$ and $\Phi_N$ are odd under the $Z_2$ symmetry. The $Z_2$ symmetry forbids the term $QH\psi^c$. Recall that  an exact CP symmetry has been imposed and all the couplings appearing in the Lagrangian, here and elsewhere, are real.  It is clear that the U(1)$_N$ symmetry is collectively broken by ${y}_\psi^i$ and $\tilde{y}_\psi^i$. This leads to the breaking of the relaxion shift symmetry and the generation of the rolling potential in Eq.~\eqref{eq:roll},\eqref{eq:mu2} with,
\begin{align}
\label{eq:LambdaH}
M&\sim\dfrac{\sqrt{y_\psi^i\tilde{y}_\psi^j(Y^{u\dagger}Y^u)_{ij}}}{4\pi}f,\\
\label{eq:rroll}
r_{\mathrm{roll}}&\sim \dfrac{4\pi \sqrt{y_\psi^k\tilde{y}_\psi^k}}{y_\psi^i\tilde{y}_\psi^j(Y^{u\dagger}Y^u)_{ij}}.
\end{align}
 The 1-loop $\Phi_N\to\Phi_N$ diagram gives the first term whereas the  $\Phi_N^2H^\dagger H$  box diagram gives the second term. For the   loop diagram generating the first term,  we have taken the the cut-off for the $\psi u^c$ loop to be the mass of the clockwork radial modes, $m_\rho \sim f$.

\subsubsection{Nelson-Barr solution to the strong CP problem}

We now show that the Lagrangian in Eq.~\eqref{eq:nb} also provides a solution to the strong CP problem. Once the relaxion stops, the phase, $\theta_N=\langle \pi_N \rangle/f \sim \phi/F$, enters the  $4\times 4$  matrix for the up sector,
\begin{equation}
M_u=
\begin{pmatrix}
(\mu_\psi)_{1\times 1} & (B)_{1\times 3}\\
(0)_{3\times 1} & (vY^u)_{3\times3}
\end{pmatrix},
\end{equation}
where,
\begin{equation}
B_i= \frac{f}{\sqrt{2}}(y^i_\psi e^{i\theta_N} +\tilde{y}^i_\psi e^{-i\theta_N}).
\end{equation}
% $\langle \pi_N|\phi\rangle\approx 1/3^N$ and thus $\pi_N/f\sim \phi/F\equiv \theta_N$, we find that the Lagrangian now has terms of the form,
The phase, $\theta_N$, is nothing but the phase of the cosine of the rolling potential  at the relaxion stopping point. Note that the bottom-left element in the above mass matrix is zero due to the absence of the $QH \psi^c$ term in the Lagrangian, which in turn is a direct consequence of the  $Z_2$ symmetry. This ensures that at tree level there is no contribution to 
$\bar{\theta}_{QCD}$ from the phase, $\theta_N$,   as,
 \begin{equation}
 \mathrm{Arg}(\det(M_u))=\mathrm{Arg}(\mu_\psi\cdot\det(vY^u))=0,
 \end{equation}
  where we use the fact that $\mu_\psi$ is real. On the other hand, for $\mu^{2}+B_{i}B_{i}^{*}\gg v^2$ we can integrate out the vector-like pair to give  the following effective $3\times3$ mass squared matrix of the SM up quark sector:
\begin{equation}\label{iout}
\left[M_{u}^{\textrm{eff}}M_{u}^{\textrm{eff}\dagger}\right]_{ij}\sim v^2 Y^{u}_{ik} Y^{u *}_{jk}-\frac{v^2Y^{u}_{ik}B_{k}^{*}B_{\ell}Y^{u *}_{j\ell}}{\mu^{2}+|B|^2}\,.
\end{equation}
For $|\vec y^{\psi} \times \vec {\tilde{y}}^{\psi}| / |\vec y^{\psi} + \vec {\tilde{y}}^{\psi}|^2\ll1\,$ there is an  ${\cal O}$(1) phase in the second term above and thus in the  unitary matrix, $V_{u_L}$, diagonalising the above matrix. This is translated into an  ${\cal O}$(1) phase in the CKM matrix $V_{CKM}=V_{u_L}^\dagger V_{d_L}$.

In general, the delicate structure of the mass matrix, $M_u$, is spoiled by radiative effects. For instance, these effects might induce a non-zero $QH \psi^c$ term, or a phase in the diagonal terms, inducing a finite $\bar{\theta}_{QCD}$. Such effects were carefully analysed in Ref.~\cite{Davidi:2017gir} where it was shown that  $\bar{\theta}_{QCD}$ is within experimental bounds provided, 
\begin{equation}
y_\psi\lesssim 10^{-2},
\end{equation}
where $\tilde{y}^i_\psi\sim {y}^i_\psi \sim y_\psi$.

Before going to the next section we would like to comment that there appears to be no obvious difficulty in extending our model along the lines of Ref.~\cite{Davidi:2018sii} to also address the SM flavour puzzle for the charged leptons and quarks. At the cost of complicating our model, this can be achieved by identifying one of the intermediate sites of the clockwork chain with the flavon  for the charged fermions and the abelian symmetry at this site with a Froggatt-Nielsen flavour symmetry.  In order not to generate a $\bar{\theta}_{QCD}$, the charge assignment of the SM fermions must be anomaly-free with respect to QCD as emphasised in Ref.~\cite{Davidi:2018sii} where an example charge assignment was also presented. We do not explore this direction further and stick to our   more minimal set-up here.
 
\begin{figure}
\noindent \begin{centering}
\hspace{-2em}\includegraphics[scale=0.53]{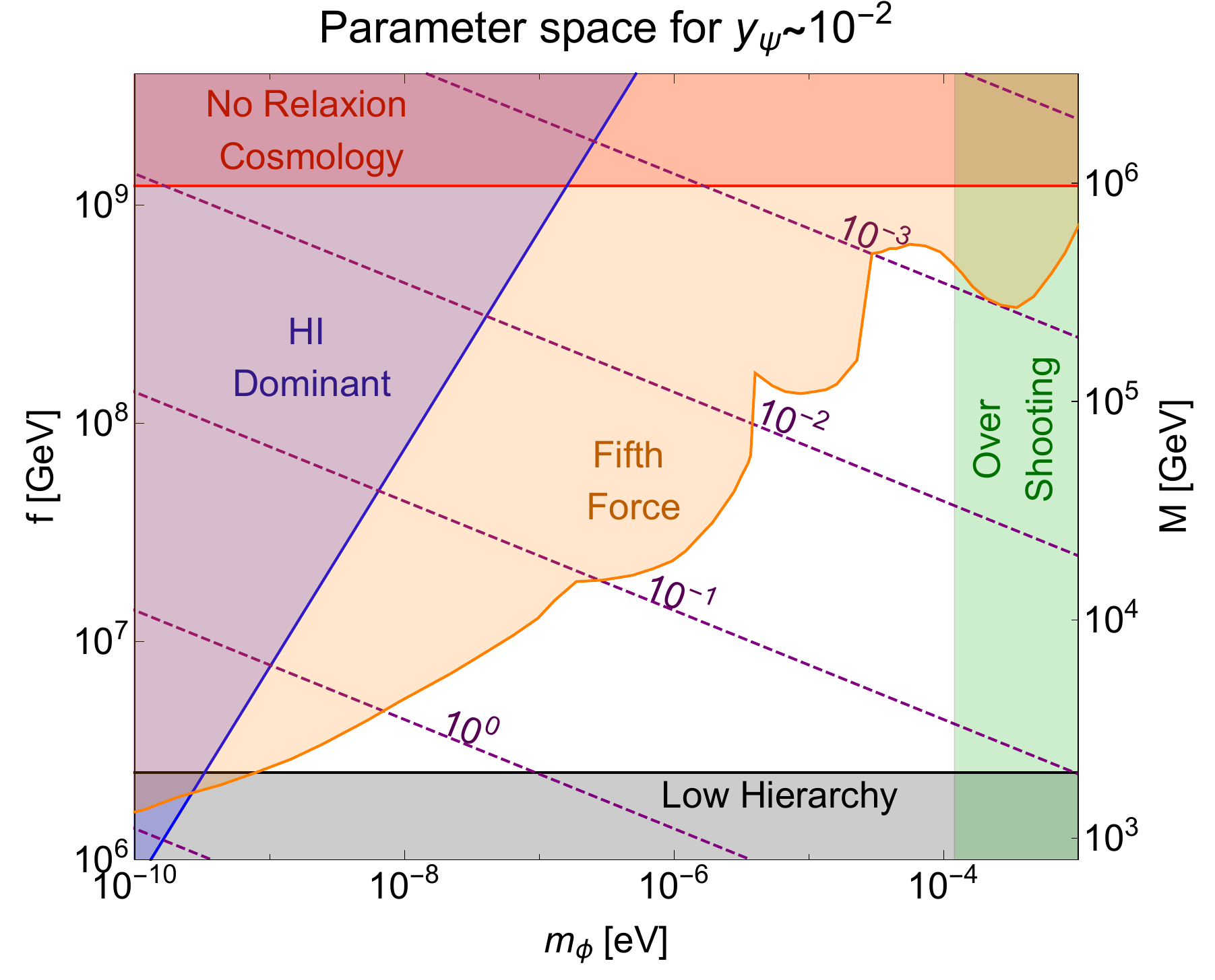}
\label{fig:plot}
\par\end{centering}
\protect\caption{The parameter space for an all-in-one relaxion in the  $(m_\phi,f)$ plane. For each value of $f$, taking the maximal value $y_\psi= 10^{-2}$, the value of the Higgs mass cut-off, $M$, is fixed as shown on the right-hand side of the frame. The red band shows the region where the cut-off exceeds the bounds imposed by cosmological requirements; blue denotes the region where the Higgs-independent contributions to the backreaction are no longer subdominant; the orange region shows the fifth force exclusions due to mixing of $\phi$ with the Higgs; green denotes the region in which the relaxion overshoots the backreaction barriers after reheating and the purple dashed lines show the value of $\tan(\phi_0/f)$  (which can also be interpreted as the required amount of tuning) needed to reproduce the correct relic density. Finally, the grey band at the bottom shows the region where the relaxion mechanism is unable to raise the Higgs cut-off beyond 2 TeV.
For more details see Sec.~\ref{sec:param}.}
\label{pspace}
\end{figure}
\section{Parameter Space}
\label{sec:param}

In this section we impose the constraints derived throughout the previous sections on the $(m_\phi,f)$ plane. The results are shown in Fig.~\ref{pspace}. We fix the $\fc/f$ ratio according to Eq.~\ref{ratio} such that each point in the plot gives the correct baryon asymmetry. First, let us consider the constraints that arise on the SRB scenario for the right-handed neutrino current introduced in Sec.~\ref{sbg}. The vertical green band shows the region that is ruled out by requiring, $m_\phi < 5 H(T_c)$, the condition in Eq.~\ref{eq:slowroll} that the relaxion does not overshoot the barriers of the backreaction potential once they reappear after reheating. Here we have taken the maximal value  $T_c= T_{sph}$ (see Sec.~\ref{sbg}). The blue shaded region corresponds to the region $\Lc^2> 16 \pi^2 v^2$ ruled out by the requirement (derived in Ref.~\cite{Graham:2015cka}) that the Higgs-dependent parts of the  backreaction potential  dominate over any Higgs-independent contribution. The dashed lines show the required  value of $\tan(\phi_0/\fc)$ to reproduce the correct dark matter density in Eq.~\ref{eq:dm}.  As explained below Eq.~\ref{eq:dm}, the extent to which $\tan \frac{\phi_0}{f}$  deviates from unity can be interpreted as a measure of the required tuning.  The orange region shows fifth force constraints that arise due to the fact that the relaxion mixes with the Higgs boson with a mixing angle (see Ref.~\cite{Flacke:2016szy}),
\begin{equation}
\sin \theta \sim \frac{\Lc^4}{\fc v m_h^2}.
\end{equation}

The rest of the constraints arise from the implementation of the Nelson-Barr mechanism in Sec.~\ref{nbr}. First of all, from Eq.~\ref{eq:LambdaH} we see that for a given value of $f$ and $y_\psi$ one can fix the value of the Higgs mass cut-off , $M$, giving us the scale on the right hand side of the frame. The red band at the top shows the region where the value of the cut-off, $M$, exceeds the upper bound imposed in Eq.~\ref{eq:relcos}. Finally the grey band at the bottom shows the region where the relaxion mechanism is unable to raise the Higgs cut-off beyond 2 TeV.

We see from Fig.~\ref{pspace} that  after all the constraints are imposed, a finite allowed region remains that  remarkably contains  the region in which tuning to obtain the correct relic density for dark matter is minimal. We will comment in the next section on how this allowed region can be probed further by  future experiments.

%
%
%\paragraph{Relaxion cosmology}
%
%We showed that the condition (Eq. \ref{eq:relcos}) must be satisfied for the relaxion rolling to be classical and for the subsequent appropriate cosmology to be realised. In our simplified case, Eq.~\ref{eq:rroll} dictates that $r_\mathrm{roll}\sim 4\pi g_\mathrm{clock}/(y_ty_\psi)\sim 10^3$ with $g_\mathrm{clock}\sim1$.\\
%
%\paragraph{Loop consistancy}
%
%Here loop consistancy refers to the need for tThis boils down to the requirement in Eq.~\ref{eq:loopcon}.\\
%
%\paragraph{Hierarchy problem}
%The relaxion mechanism solves the SM hierarchy problem up to the scale $\sim \Lambda_H$. Hence, we demand $\Lambda_H>5$TeV so that the hierarchy problem is at least significantly reduced.\\
%
%\paragraph{Overshooting}
%For successful halting of relaxion field in local minimum require $m_\phi<5H(T_c)$. \\
%
%\paragraph{Neutrino masses}
%
%\paragraph{Dark matter}
%
%

\section{Discussion and Conclusion}
%At site $0$ we have Froggart-Nielson sector, which simultaneously generates small neutrino masses and an operator suitable for spontaneous baryogenesis. At site $k$, we have a new strong-like sector (whose axion is the angular mode $\pi_k$), which generates the backreaction with analogues to axion potentials in QCD. Finally, at site $N$ we have Nelson-Barr mechanism, which introduces a CP phase and solves strong CP. Solutions to all these problems are all achieved as well as the primary feat of the relaxion: solving the hierarchy problem.
%
%The full Lagrangian of the model has the structure,
%\begin{equation}
%\begin{split}
%\mathcal{L}=&\mathcal{L}_{SM} + \mathcal{L}_{\mathrm{clock}}(\Phi_{i})\\
%&+ \mathcal{L}_{\mathrm{FN}}(\Phi_0,l_i,H,n_i)\\
%&+\mathcal{L}_{\mathrm{BR}}(\Phi_k,G',L,N)\\
%&+\mathcal{L}_{\mathrm{NB}}(\Phi_N,u^i,H,\psi).
%\end{split}
%\end{equation}
%Essentially, the non-trivial sites simultaneously generate the necessary potentials for the relaxion mechanisms and solve common BSM problems. Each site and its roll is explained in more detail in the following subsections.
%\JYR{Flow chart type diagram?}

We have presented a simultaneous solution to five BSM puzzles: the hierarchy problem, dark matter,  matter-antimatter asymmetry,  neutrino masses and the strong CP problem. While our construction is admittedly more involved than some other attempts to solve BSM puzzles in a unified way~\cite{Ballesteros:2016euj,Ballesteros:2016xej,Calibbi:2016hwq,Ema:2016ops}, this is because we also use cosmological relaxation to achieve the challenging task of solving the hierarchy problem without adding any TeV scale states that cut-off the Higgs mass divergence. Our construction has all the ingredients of a standard relaxion model -- such as a chain of clockwork scalars and a TeV-scale strong sector -- but beyond this we only make minimal modifications by adding three RH neutrinos and an up-type SU(2)$_L$ singlet vector-like quark pair.

 We utilise the fact that the standard relaxion mechanism has many interesting features built-in, such as spontaneous CPT violation during its rolling, spontaneous CP violation when it stops and oscillations about its stopping point after reheating. Motivated by previous work, we use these features to solve other BSM puzzles. The spontaneous CPT violation leads to spontaneous baryogenesis during the rolling of the relaxion after reheating; the spontaneous CP violation leads to a Nelson-Barr solution of the strong CP problem; and the relaxion oscillations generate the observed dark matter abundance. The spontaneous baryogenesis mechanism  requires that baryons and/or leptons are charged under the relaxion shift symmetry. In this work the relaxion shift symmetry is identified with a Froggatt-Nielsen symmetry, under which three new RH neutrino states (but no SM states) are charged. This satisfies the requirement of spontaneous baryogenesis while also giving an explanation for the smallness of neutrino masses for a seesaw scale below the weak scale.

Our all-in-one relaxion set-up  gives a diverse set of observational predictions. Our construction predicts the absence of a CP violating phase in the neutrino mass matrix and  GeV scale  sterile neutrinos that might not be too far from the reach of future experiments such as SHiP~\cite{Alekhin:2015byh}. The  strong CP phase in our model is non-zero  and may be detectable in future experiments. The finite allowed parameter space in Fig.~\ref{pspace} can also be probed by future improvements in  fifth force experiments. Finally, we would like to point out the interesting trade-off that exists in Fig.~\ref{pspace} between the Higgs mass cut-off scale up to which the hierarchy problem can be solved and the tuning required to reproduce the correct relic abundance of dark matter.  The least tuned regions correspond to  cut-off values smaller than 100 TeV. Top partners in a full solution to the hierarchy problem can thus be expected to be seen at these mass scales in a future 100 TeV collider.

We conclude by mentioning an interesting future direction. Our construction involves the standard relaxion mechanism, which utilises Hubble friction to stop the relaxion. It will be interesting to see if some of our ideas can be implemented in  alternative models involving particle production~\cite{Hook:2016mqo,Fonseca:2018xzp,Son:2018avk,Fonseca:2018kqf}, which are attractive because they decouple the relaxion mechanism from inflation.

\noindent
{\it Acknowledgements} We would like to thank S. Abel,  J. Scholtz  and A. Titov for very useful discussions.

\pagebreak
\bibliographystyle{unsrt}
\bibliography{citations}

\bibliographystyle{unsrt}

\end{document}